\documentclass[12pt]{article}
\usepackage{graphicx}
\usepackage{booktabs}
\usepackage{multirow}
\usepackage[numbers]{natbib}
\usepackage{centernot}
\usepackage{xcolor}
\usepackage{amssymb}% http://ctan.org/pkg/amssymb
\usepackage{pifont}% http://ctan.org/pkg/pifont
\usepackage[normalem]{ulem}

\newcommand {\bd}[1]{\mbox{\boldmath$#1$}}

\def\logit{{\mathrm{logit}}}
\def\bSig\mathbf{\Sigma}

\usepackage{chngcntr}
\counterwithout{figure}{section}

\begin{document}

\title{\bf \large Selection bias in the treatment effect for a principal stratum}

\author{ 
{\bf \small Yongming Qu* } \\ \footnotesize Department of Statistics, Data and Analytics, Eli Lilly and Company, Indianapolis, IN 46285, USA \\ \footnotesize Email: qu\_yongming@lilly.com\\
{\bf \small Stephen J. Ruberg } \\ \footnotesize Analytix Thinking, LCC, 11121 Bentgrass Court, Indianapolis, IN 46236, USA \\ \footnotesize Email: AnalytixThinking@gmail.com \\
{\bf \small Junxiang Luo } \\ \footnotesize  Moderna, Inc., 200 Technology Square, Cambridge, MA 02139, USA \\
\footnotesize Email: junxiang.luo@Modernatx.com \\ 
{\bf \small Ilya Lipkovich } \\ \footnotesize Department of Statistics, Data and Analytics, Eli Lilly and Company, Indianapolis, IN 46285, USA \\ \footnotesize Email: ilya.lipkovich@lilly.com \\
}
\date{\small \em \today}
\maketitle
\noindent
{\footnotesize *Correspondence: Yongming Qu, Department of Statistics, Data and Analytics, Eli Lilly and Company, Lilly Corporate Center, Indianapolis, IN 46285, U.S.A. Email: qu\_yongming@lilly.com.}

\label{firstpage}
\newpage
%  put the summary for your paper here

\begin{abstract}
Estimation of treatment effect for principal strata has been studied for more than two decades. Existing research exclusively focuses on the estimation, but there is little research on forming and testing hypotheses for principal stratification-based estimands. In this brief report, we discuss a phenomenon in which the true treatment effect for a principal stratum may not equal zero even if the two treatments have the same effect at patient level which implies an equal average treatment effect for the principal stratum. We explain this phenomenon from the perspective of selection bias. This is an important finding and deserves attention when using and interpreting results based on principal stratification. There is a need to further study how to form the null hypothesis for estimands for a principal stratum.
 
\end{abstract}

%  Please place your key words in alphabetical order, separated
%  by semicolons, with the first letter of the first word capitalized,
%  and a period at the end of the list.
%

\maketitle
\newpage

%\begin{abstract}
%\noindent
%{\small SUMMARY.  }
%\end{abstract}

%\noindent 
%{\small
%{\bf KEY WORDS} Information gain, Kullback-Leibler, Likelihood reduction factor, Treatment effect, Surrogate marker.
%}
\section{Introduction} \label{sec:intro}
A principal stratum is a subset of subjects defined by potential outcomes of postbaseline variable(s). Estimation of the treatment effect for a principal stratum was first introduced in the late 1990s and early 2000s \cite{angrist1996identification,imbens1997bayesian,frangakis2002principal}. The newly published ICH E9 Addendum (R1) listed principal stratification as one of the key strategies for handing intercurrent events and defining population when forming estimands \cite{international2021harmonised}. There are many other clinically meaningful scenarios in which a principal stratum can be of great importance, whether it is a primary, secondary, or supplemental estimand. In parallel studies, identification of the principal strata is always challenging. Many efforts, often with additional assumptions, have been proposed to estimate the treatment effect for principal strata in parallel studies. Lipkovich et al. \cite{lipkovich2021using} provided a comprehensive review of all current principal stratification-based methods.  

There are generally two types of principal strata: (1) the principal strata based on one or more postbaseline variables under one treatment (e.g., patients who can adhere to the experiment treatment), and (2) the principal strata based on one or more postbaseline variables under multiple (two or more) treatments (e.g., patients who can adhere to both treatments). In application, either type of principal strata can be of interest. For example, Qu et al. \cite{qu2020general,qu2021implementation} suggested the former may be useful for placebo-controlled studies and the latter may be useful for active-comparator studies. More recently, Bornkamp et al. \cite{bornkamp2021principal} provides more discussions on the utilization of different principal strata in clinical trials. However, given the null scenario where the true treatment effect is zero for each patient in the entire study population, it remains an open question whether the true treatment effect for the defined principal strata also equals zero. This has not been discussed except briefly by Luo et al. \citep{luo2021estimating},  
%However, whether the true treatment effect (estimand) for these principal strata equals zero under the null scenario of the experimental and the comparator treatments having the exact same effect (both efficacy and safety) at the patient level has not been discussed, except briefly by Luo et al. \citep{luo2021estimating}. 
which may be attributable to people's thinking that the treatment effect should always equal zero for any principal stratum under the null scenario. However, recently we found that it is not necessarily true.

In this report, we will describe how the true treatment effect in a principal stratum may not equal zero even under the null scenario, showing more research is needed in forming hypothesis testing when using principal stratification-based estimands.

\section{Methods}
\label{sec:methods}
We use the notation in Qu et al. \cite{qu2020general} and adherence status as the principal stratification variable, but the argument can be applied to any principal stratification variable. 

Let $(\bd X_{j}, T_j, \bd Z_{j}, Y_{j}, A_{j})$ denote the data for subject $j \; (1 \le j \le n)$, where $\bd X_{j}$ is a vector of baseline covariates, $T_j$ is the assigned treatment, $\bd Z_{j} = (Z_{j}^{(1)}, Z_{j}^{(2)}, \ldots, Z_{j}^{(K-1)})'$ is a vector of intermediate repeated measurements, $Y_j$ is the outcome of interest, and $A_{j}$ is the indicator variable for whether a patient is adherent to the assigned treatment. Note $\bd Z_{j}$ can be intermediate measurements of the same variable as $Y$, or intermediate outcomes of other ancillary variables, or include both.
We use ``$(t)$" following the variable name to denote the potential outcome under the hypothetical treatment $t \; (t=0,1)$ \cite{holland1986statistics}. For example, $Y_{j}(t)$ denotes the potential outcome for subject $j$ if taking treatment $t$. Generally, $Y_{j}(T_j)$ can be observed but $Y_{j}(1-T_j)$ cannot be observed in parallel studies. To simplify notation, we may drop the subscript $j$ from the random variables if it is not needed for clarity. 

%Let $Y_j$ be a random variable of the final outcome, $X_j$ be a baseline covariate vector, $Z_j$ be a vector of intermediate post baseline measurements, $T_j$ be a treatment indicator ($T_j = 0$ for the control group and $T_j = 1$ for the experimental treatment group), and $A_j$ denote the adherence status for the assigned treatment for the planned duration of the trial, where $A=1$ means that a patient taking assigned treatment completes the trial and observes the primary endpoint. The vector $Z$ may include intermediate efficacy and safety outcomes that can affect the probability of adherence (or non-adherence) due to different reasons (e.g., lock of efficacy or adverse events). Since $Y$, $Z$, and $A$ denote the variables under actual assigned treatment, we use $Y(T)$, $Z(T)$, and $A(T)$ to denote their potential outcomes under treatment $T$, originally introduced by \cite{holland1986statistics}. 

As an illustration, we consider two principal strata
\begin{list}{$\bullet$}{\leftmargin=0cm \itemindent=1.5cm}
    \item Patients who can adhere to the experimental treatment: $S_{*+}=\{j: A_{j}(1)=1\}$
    \item Patients who can adhere to both treatments: $S_{++}=\{j: A_{j}(0)=1, A_{j}(1)=1\}$
\end{list}

Let us first consider a simple null scenario in which there is no treatment difference for all efficacy and safety parameters at the patient level. Under such a null scenario, $\{\bd X_{j}, \bd Z_{j}(t), Y_{j}(t), A_{j}(t)\}$ have the same distribution for $t=0, 1$. By symmetry, the true treatment effect for the principal stratum $S_{++}$ is given by
\begin{eqnarray}
\mu_{d, ++} &:=& E\{Y(1)-Y(0)|A(0)=1,A(1)=1\} \nonumber\\ &=& E\{Y(1)|A(0)=1, A(1)=1\} - E\{Y(0)|A(0)=1, A(1)=1\}  \nonumber\\ &\equiv& 0.
\end{eqnarray}
However, 
\begin{eqnarray}
\mu_{d, *+} &:=& E[Y(1)-Y(0)|A(1)=1] \not\equiv 0. 
\end{eqnarray}
A sufficient condition for $\mu_{d,*+} = 0$ under the null scenario is
\begin{equation} \label{eq:sufficient}
\{Y(1)-Y(0)\} \perp A(1) | X.
\end{equation}
This can be seen by
\begin{eqnarray}
\mu_{d, *+} &=& E[E\{Y(1)-Y(0)|A(1)=1, X\}|A(1)=1]
\nonumber\\&=&
E[E\{Y(1)-Y(0)|X\}|A(1)=1] \quad\quad\quad\quad \mbox{by condition (\ref{eq:sufficient})} 
\nonumber\\&=&
0.
\end{eqnarray}

Let us consider a data generation model used in the simulation in Qu et al. \cite{qu2020general}. The baseline and hypothetical outcomes are generated by
\begin{equation} \label{eq:model_x}
X_j \sim NID(\mu_x, \sigma_x^2),
\end{equation}
\begin{equation} \label{eq:model_z}
Z_{j}^{(k)}(t) = \alpha_{0k} + \alpha_{1k} X_j + \alpha_{2k} t + \eta_{j}^{(k)}(t), \quad  1 \le k \le 3, 
\end{equation}
and
\begin{equation} \label{eq:model_y}
Y_j(t) = \beta_0 + \beta_{1} X_j + \beta_{2} t + \sum_{k=1}^3 \beta_{3k} Z_{j}^{(k)}(t) + \epsilon_j(t), 
\end{equation}
where {\em NID} means normally independently distributed, $k$ indicates the time point for repeated measures for the intermediate outcome, 
$\left\{\eta_{j}^{(k)}(t): j=1,2,\ldots,n; t=0,1 \right\} \sim NID(0, \sigma_\eta^2)$ and $\left\{\epsilon_j(t): j=1,2,\ldots,n; t=0,1 \right\} \sim NID(0, \sigma_\epsilon^2)$,
and  $\eta_{j}^{(k)}(t)$'s and $\epsilon_j(t)$'s are independent. The treatment code $T_j$ is generated independently from a Bernoulli distribution with probability of 0.5.

The adherence status right after time point $k \; (1\le k \le 3)$ is generated from a logistic model
\begin{eqnarray} \label{eq:model_A}
\logit\{\Pr(A_{j}^{(k)}=1|A_j^{(k-1)}=1, X_j, Z_{j}^{(k)})\} = \gamma_0 +  \gamma_1 X_j +  \gamma_{3k} Z_{j}^{(k)}, 
\end{eqnarray}
where $\logit(p)=\log(p/(1-p))$, and by convention we set $A_j^{(0)}=1$. Let $A_j = \prod_{k=1}^3 A_j^{(k-1)}$. 

Based on the Equation (B.5) in Appendix B of Qu et al. \cite{qu2020general}, under the null scenario of no treatment difference ($\alpha_{2k}=0$ and $\beta_2=0$), the treatment effect for the principal stratum of $S_{*+}=\{A(1)=1\}$ is given by
\begin{eqnarray}
\mu_{d,*+}
= 
\frac{
\int\int\int\int
\frac{\left( 
\sum_{k=1}^3 \beta_{3k} \xi_{k} 
\right) f(x|\mu_x,\sigma_x^2) f(\xi_1|0, \sigma_\eta^2) f(\xi_2|0, \sigma_\eta^2) f(\xi_3|0, \sigma_\eta^2)}
{ 
\prod_{k=1}^3 \left[ 1 + \exp\{-(\gamma_0+\gamma_{3k}\alpha_0)-(\gamma_1+\gamma_{3k}\alpha_{1k}) x - \gamma_{3k} \xi_k\} \right]
}
dx d\xi_1 d\xi_1 d\xi_3
}
{
\int\int\int\int
\frac{f(x|\mu_x,\sigma_x^2) f(\xi_1|0, \sigma_\eta^2) f(\xi_2|0, \sigma_\eta^2) f(\xi_3|0, \sigma_\eta^2) }
{\prod_{k=1}^3 [1 + \exp\{-(\gamma_0+\gamma_{3k}\alpha_0)-(\gamma_1+\gamma_{3k}\alpha_{1k}) x - \gamma_{3k} \xi_k\}] }
dx d\xi_1 d\xi_1 d\xi_3
}.
\label{eq:mu_2}
\end{eqnarray}
It is clear that $\mu_{d,*+} \ne 0$ when $\beta_{3k} \ne 0$ and $\gamma_{3k} \ne 0$. When $\beta_{3k} = 0$ or $\gamma_{3k} = 0$, which implies the sufficient condition (\ref{eq:sufficient}) is satisfied, $\mu_{d,*+}=0$. 

The root cause of the nonzero true treatment effect for $S_{*+}$ under the null scenario is selection bias. In the illustrative example, the principal stratum $S_{*+}$ is defined by the variable $A(1)$ which is correlated with $Y(1)$, but not  with $Y(0)$, conditional on $X$. This causes asymmetric selection biases between the average responses  in $Y(0)$ and $Y(1)$ for $S_{*+}$. With the sufficient condition (\ref{eq:sufficient}) that guarantees $E\{Y(1)-Y(0)|S_{d,*+}\} = 0$ under the null treatment effect, the outcomes $Y(1)$ and $Y(0)$ are not influenced by $A(1)$ given $X$, which effectively eliminates the difference in selection bias between the two treatments.

\section{Summary}
In summary, the treatment effect for a principal stratum may not equal zero even when treatment has no effect (in both efficacy and safety) at all. While the treatment effect for a principal stratum defined by the same condition for potential outcomes under both treatments (e.g., $S_{++}$ in this article) always equals zero, due to selection bias, the treatment effect for a principal stratum defined by potential outcomes under one treatment group (e.g., $S_{*+}$) generally does not equal zero, under the null scenario. This can be a challenge when using some principal strata to define an estimand, as it may lead to a nonzero treatment effect for an ineffective treatment. 

We identify one sufficient condition (\ref{eq:sufficient}) that guarantees $E\{Y(1)-Y(0)|S_{d,*+}\} = 0$ under the null treatment effect. However, this condition is rather restrictive. It means the treatment effect for $S_{d,*+}$ can be estimated by only modeling the outcome through baseline covariates or the probability of belonging to the principal stratum can be modeled through baseline covariates, which may be too simplistic. 

The principal stratum defined by the same condition for potential outcomes under both treatments can always guarantee the treatment effect is zero under the null scenario of no treatment effect, so it can preserve the type 1 error for superiority studies; however, for non-inferiority studies, the situation becomes complex. The potential asymmetric selection bias in the treatment effect for a principal stratum naturally leads to the need for using both the outcome measurement AND the principal stratum variable(s) to form the hypothesis for principal stratification-based estimands, which has not been discussed in any literature.

One way to make the estimand be zero under the null scenario of no treatment effect is to adjust for the selection bias. For example, one can estimate the treatment effect for a principal stratum for two samples from a random split of patients in the control treatment; such a process can be repeated many times and the average treatment effect is the treatment effect for the principal stratum under the null hypothesis. This approach could be limited by its assumption that treatment and control have the same efficacy and safety under the null scenario.  If a treatment has no efficacy in the outcome $Y$ but has an effect on the potential outcome used to form the principal stratum, such an approach may not correctly estimate the treatment effect for the principal stratum under such a \emph{partial} null scenario. Until now, most critiques for the principal stratification approach are about the strong assumptions in estimation. This article shows that forming a hypothesis for the treatment effect for a principal stratum may be challenging and requires further research.

\section*{Acknowledgements}
We would like to thank Yu Du for his scientific review of this article and useful comments and Angela Lorio for an editorial review of this article.

\bibliographystyle{vancouver}
\bibliography{references}

\end{document}